\documentclass[a4paper]{jpconf}
\usepackage{graphicx}
\begin{document}
\title{Electroweak boson production in heavy-ion collisions with CMS.}

\author{Michael Gardner on behalf of the CMS Collaboration}

\address{University of California, Davis, USA}

\ead{michael.david.gardner@cern.ch}

\begin{abstract}
The Compact Muon Solenoid (CMS) is fully equipped to measure leptonic decays of electroweak probes in the high multiplicity environment of nucleus-nucleus collisions. Electroweak boson production is an important benchmark process at hadron colliders. Precise measurements of W and Z production in heavy-ion collisions can help to constrain nuclear PDFs as well as serve as a standard candle of the initial state in PbPb collisions at the LHC energies. The inclusive and differential measurements of the Z boson yield in the muon and electron decay channel will be presented, establishing that no modification is observed with respect to pp collisions, scaled by the number of incoherent nucleon-nucleon collisions.
\end{abstract}

\section{Introduction}
$W$ and $Z$ bosons were first observed by the UA1 and UA2 experiments at CERN thirty years ago in proton-antiproton collisions at $\sqrt{s} = 540$ GeV \cite{UA1_Collaboration}. Since then a succession of electron-positron, proton-proton (pp) and
proton-antiproton collider experiments have been used to measure their properties, such as mass, decay width and production cross sections at a variety of center-of-mass energies. Now for the first time, due to the large center-of-mass energy and luminosities offered at the LHC, the $Z$ and $W$ boson production can be studied in nucleus-nucleus collisions. Based on the first lead-lead (PbPb) collisions corresponding to 7.2 $\mu$b$^{-1}$ integrated luminosity, the CMS collaboration reported first results on the $Z \rightarrow \mu \mu$ \cite{CMS_Z} and $W \rightarrow \mu \nu$ \cite{CMS_W} processes, showing electroweak bosons are essentially unmodified by the hot and dense medium created in heavy-ion collisions, often referred to as the quark-gluon plasma (QGP).

$Z$-boson production is reported in this proceedings with the second set of PbPb collisions that occurred in late 2011 and accounts for 150~$\mu$b$^{-1}$ integrated luminosity. In this document, the second set of pp data at the same energy in the center-of-mass, recorded in 2013, is used as a reference in order to calculate the nuclear modification factor, $R_{\rm AA}$. The increases in statistics of both the pp and PbPb samples allow more detailed study of the transverse momentum ($p_{T}$), rapidity ($y$) and collision centrality dependencies of the $Z$-boson yields. The dimuon channel is analyzed with an improved reconstruction algorithm with respect to Ref. \cite{CMS-PAS-HIN-12-008}, and electrons are used for the first time to reconstruct the Z decays in the heavy-ion environment in CMS \cite{CMS_detector}.

$Z$ bosons, once produced, decay within the medium with a typical lifetime of 0.1 fm/c. Since leptons pass freely through the strongly-interacting medium, dileptons from $Z$ bosons can serve as a reference to the processes expected to be heavily modified in the QGP, such as quarkonium production, or the production of an opposite-side jet in $Z$+jet processes~\cite{NoInteractingZ}. However, in heavy-ion collisions, $Z$-boson production can be slightly affected by initial-state effects. This modification is expected to be about 3\% from the different mix of protons and neutrons in AA collisions (isospin effect)~\cite{Isospin}. Energy loss and multiple scattering of the initial partons could produce modifications of the order of 3\%~\cite{NRJloss}. In addition, the parton distribution functions can be depleted in nuclei, which could modify the $Z$-boson yield by at most 20\% \cite{Isospin}.

\section{Lepton reconstruction and Signal Extraction}

\subsection{Muon reconstruction in CMS}
With respect to the previous analysis reported in Ref.~\cite{CMS-PAS-HIN-12-008}, a new muon offline reconstruction algorithm is used for the PbPb data, with increased efficiency via multiple iterations in the pattern recognition step. The single-muon reconstruction efficiency is raised from $\simeq 85$\% to $\simeq 98$\%, for muons from $Z$ decays, reaching the level of the algorithm used for pp collisions. Background muons from cosmic rays and heavy-quark semileptonic decays are rejected by requiring a transverse (longitudinal) distance of closest approach of less than 0.2 (5.0) mm from the measured vertex. These selection criteria further reject about 2\% of real muons for both pp and PbPb collisions. 

\subsection{Electron reconstruction in CMS}
The electron reconstruction uses information from the pixel and strip tracker, and the electromagnetic calorimeter (ECAL). Electrons produced in pp or PbPb collisions traversing the silicon layers of the tracker can emit bremsstrahlung photons and deposit energy in the ECAL with a significant spread in the azimuthal direction. An ECAL clustering algorithm and in particular the building of so-called \emph{superclusters}, was used for collecting and estimating the proper energy of photons in the heavy-ion environment as in Ref. \cite{Chatrchyan:2012vq}. Pixels facing these superclusters are looked for, in order to initiate the building of trajectories in the inner tracker. A dedicated algorithm which takes into account the bremsstrahlung emission is used \cite{Reconstruct_GSF}.

The reconstruction algorithm results in a reconstruction efficiency of ~85\% for electrons from $Z$ decays. Further selection criteria reduce this efficiency by about 10\%. For the pp sample, the standard algorithms and identification criteria presented in Ref.~\cite{CMS:2010bta} are used, resulting in a reconstruction efficiency of $98\%$ and a further reduction of $5\%$ due to selection criteria.

\subsection{Signal Extraction}
\begin{figure}[hbtp]
  \begin{center}
     \includegraphics[angle=0, width=0.32\textwidth]{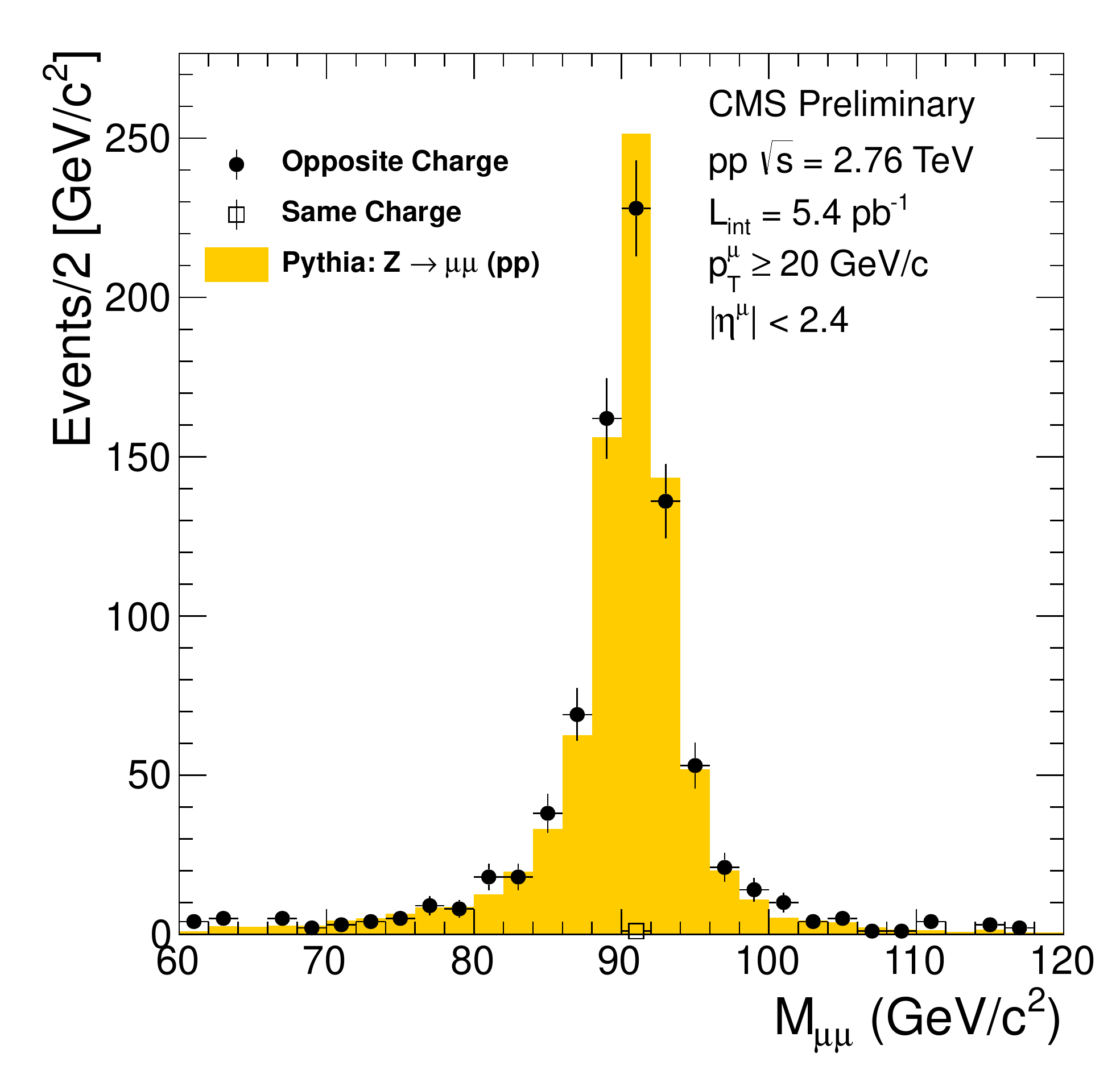}
     \includegraphics[angle=0, width=0.32\textwidth]{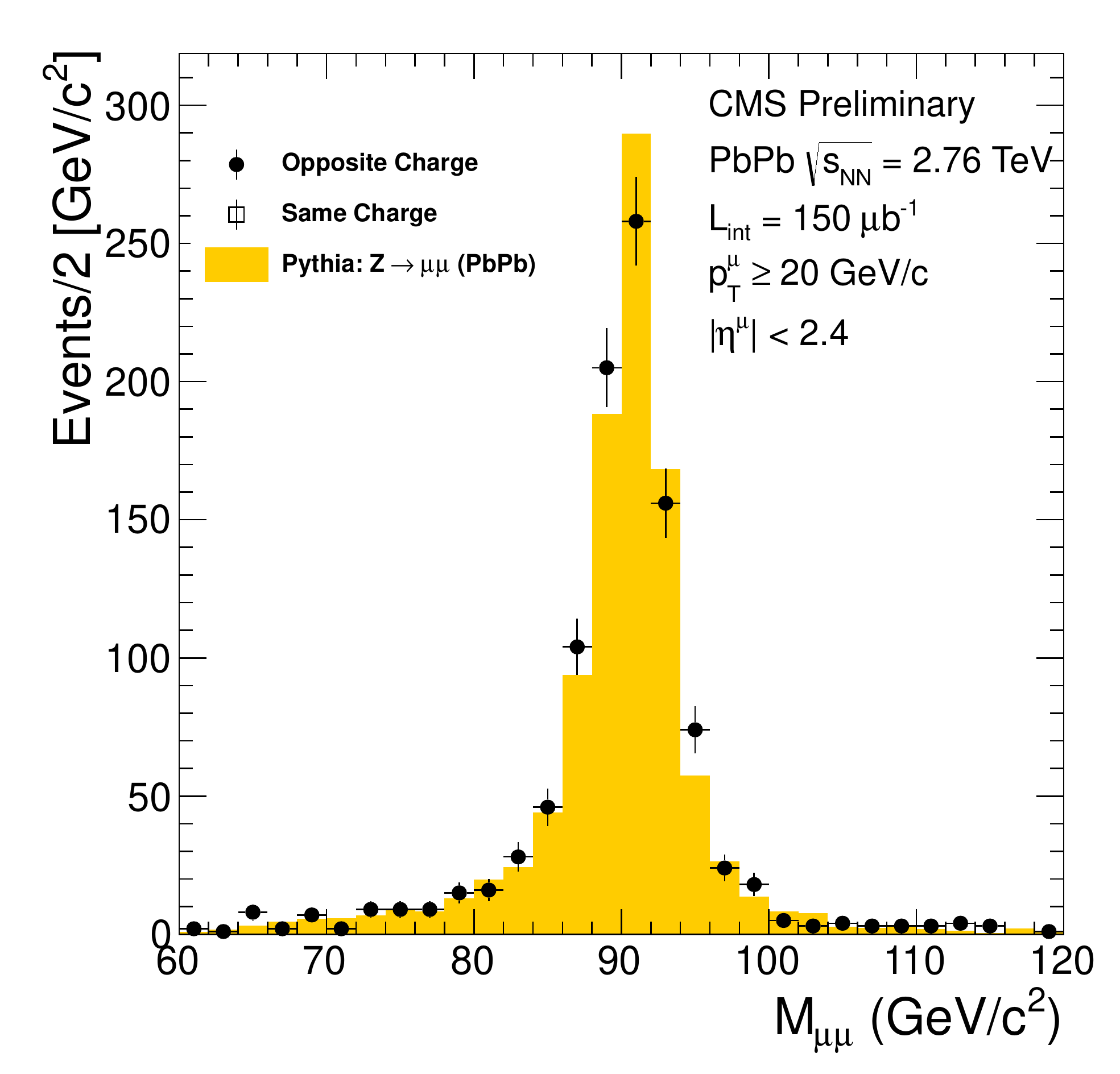}      
     \caption{Dimuon invariant mass spectra for muons with $|\eta^{\mu}| < $ 2.4 and $p_{T}^\mu > $ 20 GeV/c, and dimuons with $|y| < $ 2.0 in pp (left) and in PbPb (right) collisions. Full black circles are opposite-charge pairs, open black squares are same-charge pairs. Superimposed is the MC simulation from {\sc pythia} pp $\rightarrow Z \rightarrow \mu^{+}\mu^{-}$ embedded in {\sc hydjet} for the PbPb case.}
    \label{InvMassMuon}
  \end{center}
\end{figure}

\begin{figure}[hbtp]
  \begin{center}
    \includegraphics[angle=0, width=0.32\textwidth]{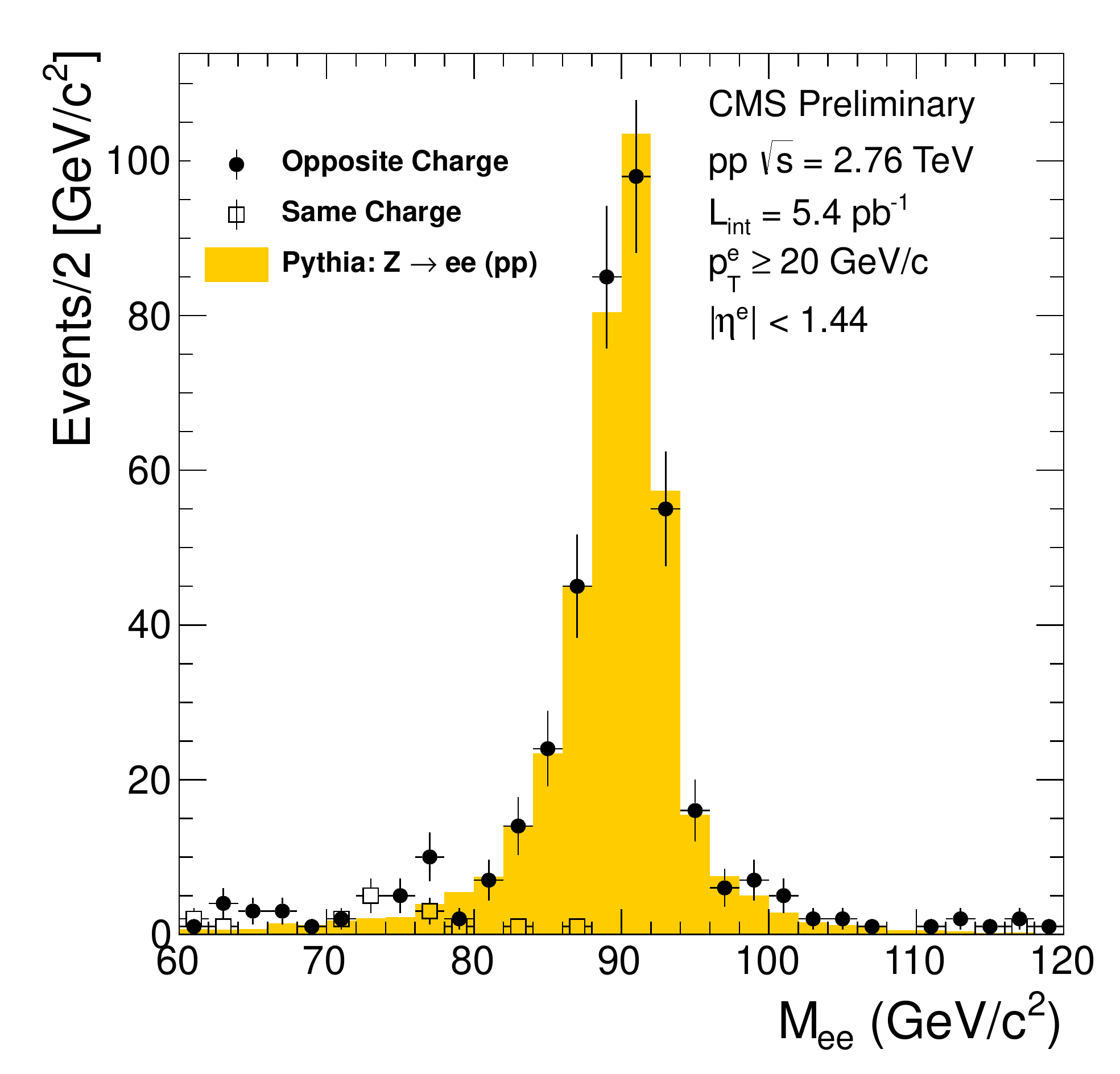}
    \includegraphics[angle=0, width=0.32\textwidth]{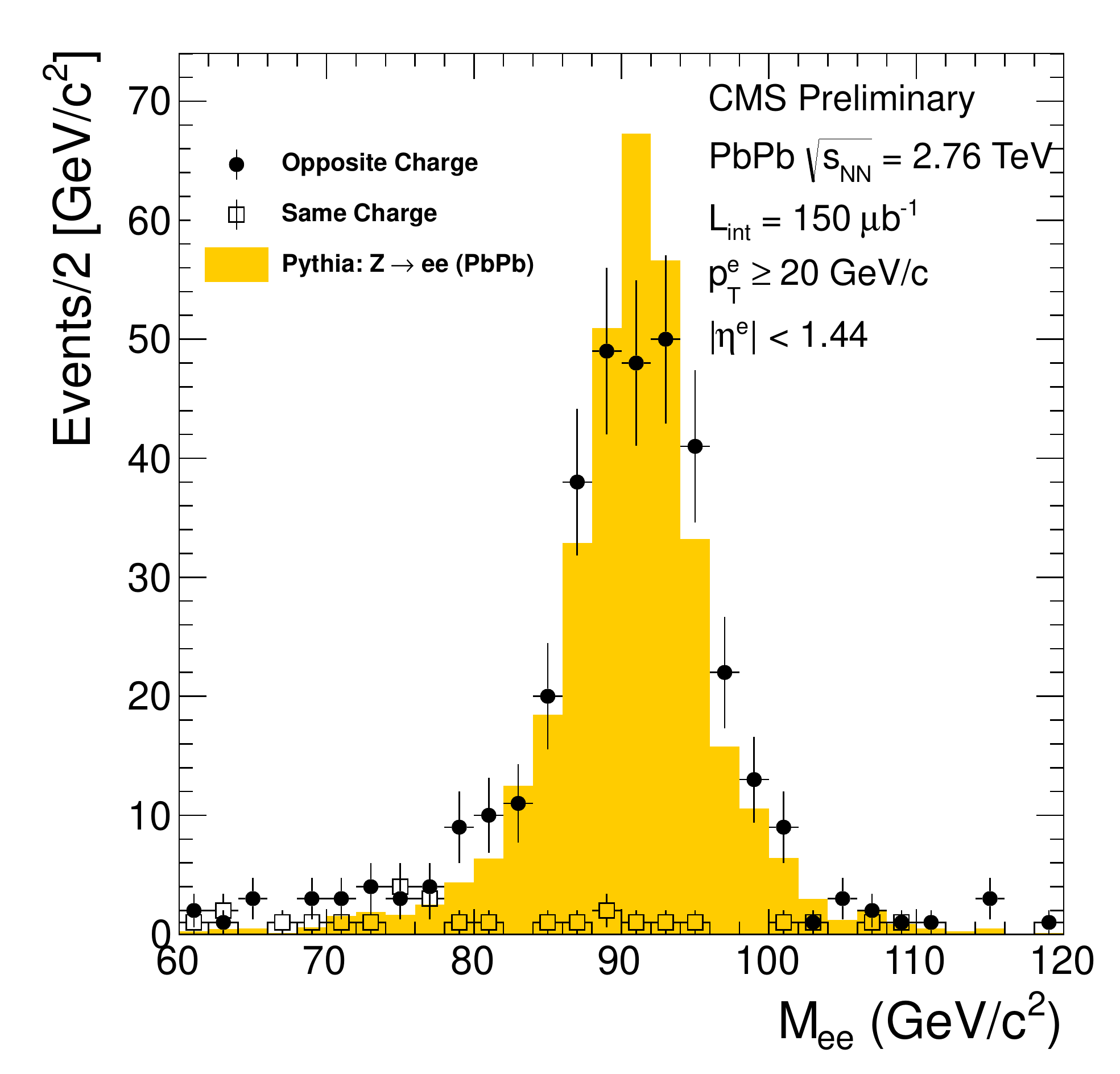}
    \caption{Dielectron invariant mass spectra for electrons with $|\eta^{e}| <$ 1.44 and $p_{T}^{e} > $ 20 GeV/c, and dielectrons with $|y| < $ 1.44 in pp (left) and in PbPb (right) collisions.}
    \label{InvMassElectron}
  \end{center}
\end{figure}

Figures \ref{InvMassMuon} and \ref{InvMassElectron} show the dimuon and dielectron invariant mass spectra in the 60--120~GeV/$c^2$ mass range after applying selection criteria in pp and PbPb collisions. Only leptons with $p_{T} > 20$ GeV/$c$ and $|\eta| < $ 2.4 and 1.44 for the muons and the electrons, respectively, are considered. The dimuon rapidity is limited to $|y|<2$, while the dielectron rapidity is kept in the $|y|<1.44$ range. Overlaid histograms are from the Monte Carlo (MC) simulation.

In the 60--120~GeV/$c^2$ mass range of the PbPb sample, 1022 events are found with opposite-charge (OC, black solid circles) muons and not a single same charge (SC, black open squares) muon pair, while the pp sample has 830 OC and 1 SC pair. In the restricted dielectron rapidity range, 328 (388) OC electrons are measured in the PbPb (pp) data, with 27 (17) SC pairs resulting from higher fake and charge mis-identification rates. The SC pairs provide a measure of uncorrelated combinatorial background: negligible (1/1000 level) in the muon channel, and $~$8\% (4)\% in the electron channel. The number of $Z$ candidates is taken as the OC--SC difference. Other sources of background are found to be negligible and covered by systematic uncertainties.

\section{Results}

Based on pp and PbPb $Z$-boson yields at the same energy in the center-of-mass, we compute the nuclear modification factors $R_{\rm AA}$ for both electron and muon channels as a function of the $Z$-boson $p_{T}$ and rapidity, and event centrality, as the following: $R_{\rm AA} = N_{\rm AA} / T_{\rm AA} \times \sigma_{pp}$, where $N_{\rm AA}$ are the Z boson yields per event estimated from PbPb collisions corrected for acceptance and efficiency measured in simulation, and $\sigma_{pp}$ the differential cross sections estimated from pp collisions by dividing the corrected Z-boson yields by the integrated luminosity. The nuclear overlap function $T_{\rm AA}$, computed with a Glauber model  \cite{Miller:2007ri} is proportional to the number of elementary nucleon-nucleon binary collisions $N_{\rm coll} = T_{\rm AA} \times \sigma_{NN}$ where the latter is an assumed inelastic nucleon-nucleon cross section that cancels to first order in $T_{\rm AA}$.

\begin{figure}[hbtp]
  \begin{center}
    \includegraphics[angle=0, width=0.32\textwidth]{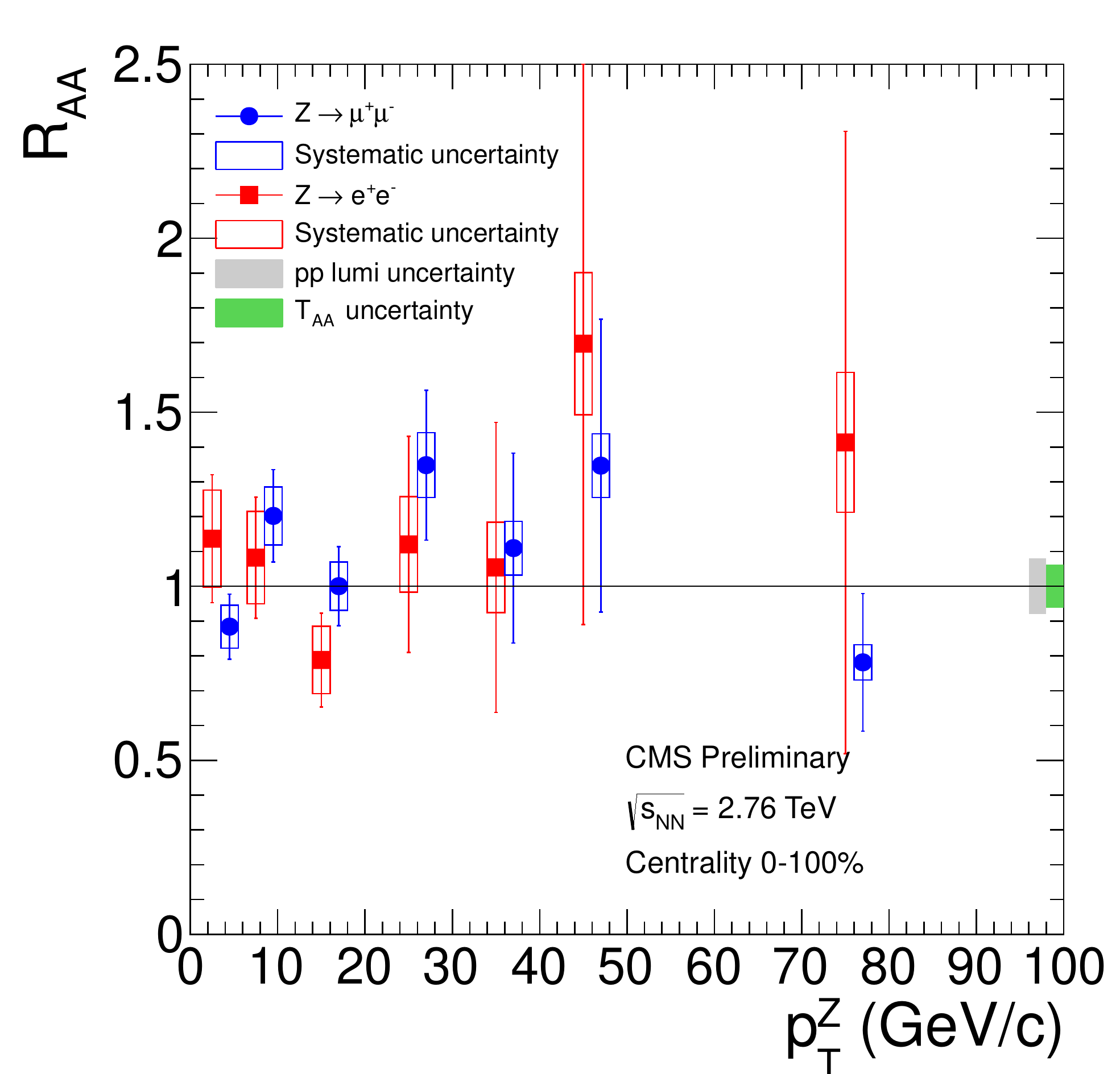}
    \includegraphics[angle=0, width=0.32\textwidth]{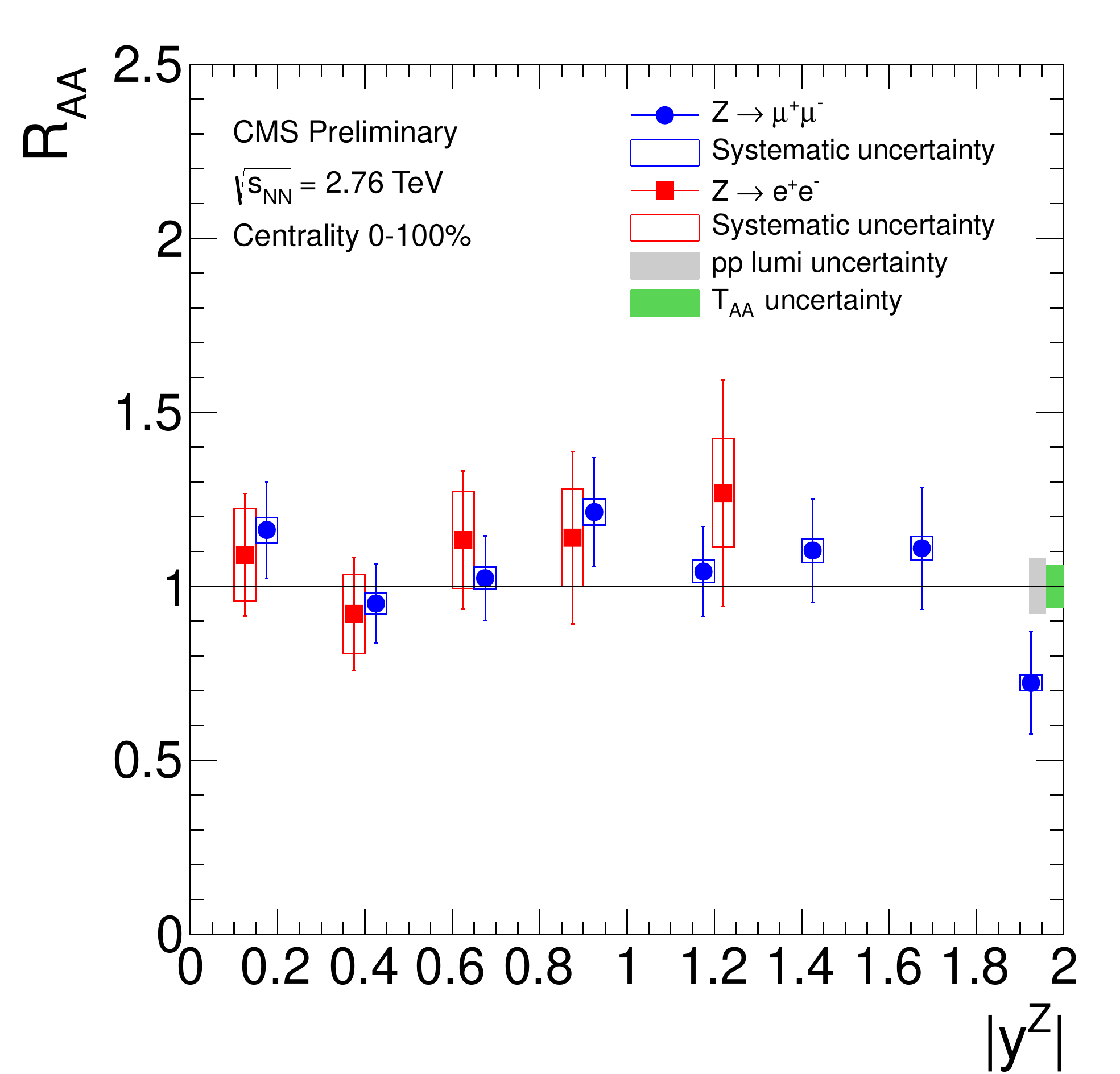}
    \includegraphics[angle=0, width=0.32\textwidth]{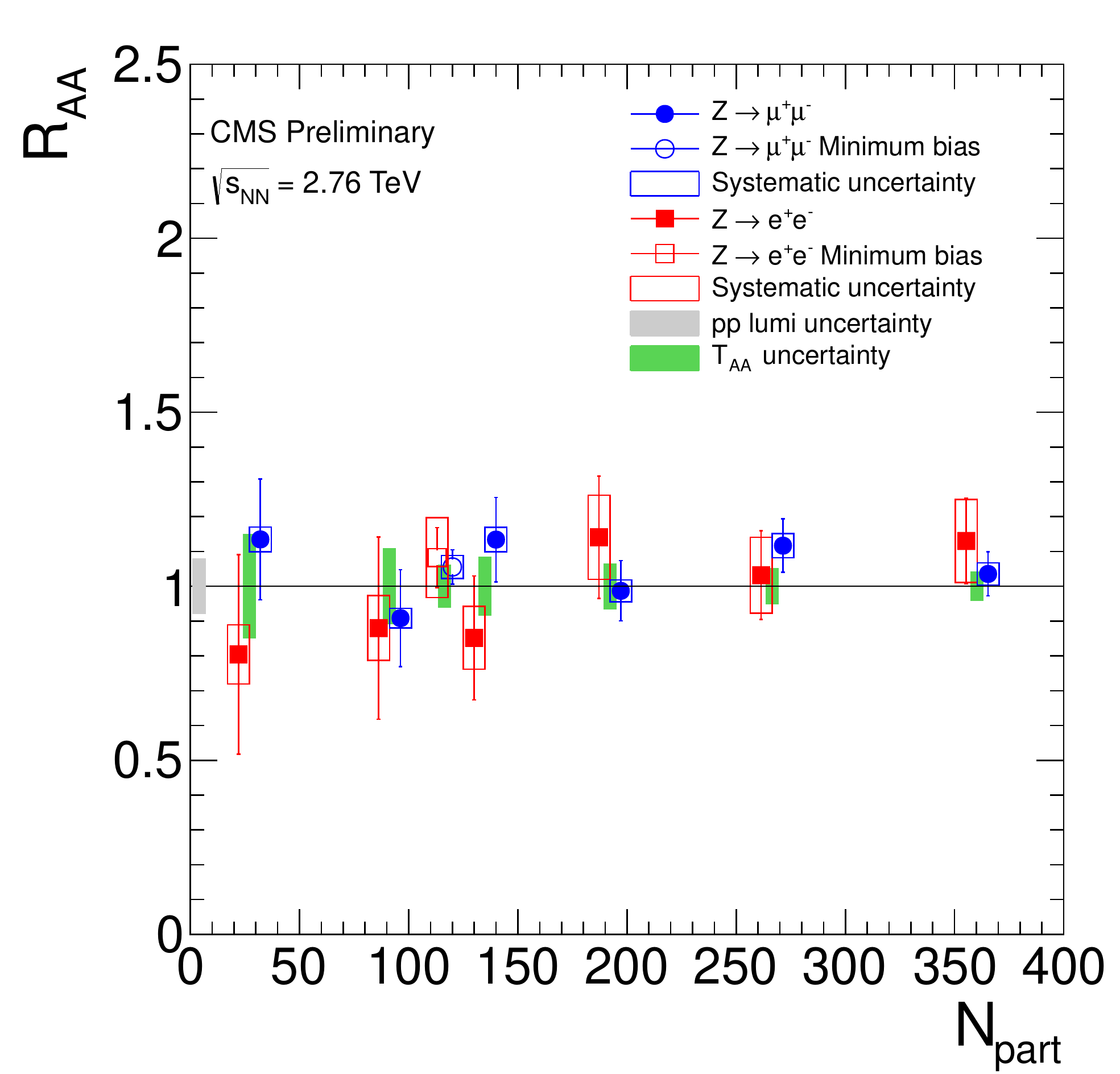}
 \caption{Nuclear modification factor ($R_{\rm AA}$) extracted from PbPb and pp collisions at $\sqrt{s_{NN}} = 2.76$ TeV for $Z \rightarrow e^+ e^-$ and $Z \rightarrow \mu^+ \mu^-$ as a function of $Z$-boson $p_{T}$ (left), rapidity (center) and number of participants in the collision (right).}
  \label{fig:RAA}
  \end{center}
\end{figure}

The data are divided in different independent ranges: 6 in event centrality, 8 (5) in dimuon (dielectron) rapidity, and 7 in Z-boson $p_{T}$. Figure \ref{fig:RAA} shows the obtained $R_{\rm AA}$ as a function of the $Z$-boson $p_{T}$, rapidity and event centrality (translated to the average number of participants ($N_{\rm part}$), through the Glauber model mentioned above). The $T_{\rm AA}$ uncertainties, displayed as vertical green bands are common to all points when centrality-integrated, and to the dielectron and dimuon points when centrality-binned. The pp luminosity uncertainty, common to all points is displayed as a vertical grey band.

Within our statistical and systematic uncertainties (dominated by uncertainty in the efficiency from a data-driven Tag and Probe method)  the nuclear modification factors in both the electron and the muon channels as a function of $p_{T}$, rapidity and event centrality are compatible with unity and no strong deviations are observed. The integrated $R_{AA}$ values are $1.06 \pm 0.05 (stat) \pm 0.11 (syst)$ in the dimuon channel and $1.08 \pm 0.09 (stat) \pm 0.14 (syst)$ for dielectrons, and the differential values are shown in Figure \ref{fig:RAA}.

\section{Conclusions}

The $Z$-boson $R_{AA}$ has been measured from PbPb and pp collisions at $\sqrt{s_{\rm NN}} = 2.76$~TeV. This was done as a function of $Z$-boson transverse momentum, rapidity and event centrality in both the dimuon and dielectron decay channels with 150~$\mu$b$^{-1}$ integrated luminosity. No significant deviation is observed, underlying that within uncertainties, no nuclear effects are observed.


\section*{References}
\begin{thebibliography}{10000}
\bibitem{UA1_Collaboration} UA1 Collaboration, ``Experimental Observation of Lepton Pairs of Invariant Mass Around $95-GeV/c^{2}$ at the CERN SPS Collider", {\it Phys.Lett. B} {\bf 126} (1983) 398-410.
\bibitem{CMS_Z} CMS Collaboration, ``Study of Z boson production in PbPb collisions at nucleon-nucleon center of mass energy =
2.76 TeV", {\it Phys.Rev.Lett.} {\bf 106} (2011) 212301, arXiv:1102.5435 [nucl-ex].
\bibitem{CMS_W} CMS Collaboration, ``Study of W boson production in PbPb and pp collisions at $\sqrt{s_{NN}} = 2.76$ TeV”, CMS, {\it Phys.Lett. B} {\bf 715} (2012) 66, arXiv:1205.6334 [nucl-ex].
\bibitem{NoInteractingZ} V.Kartvelishvili, R.Kvatadze and R.Shanidze, ``On $Z$ and $Z$+jet production in heavy ion collisions", {\it Phys.Lett. B} {\bf 356} (1995) 589, arXiv:hep-ph/9505418.
\bibitem{Isospin} H.Paukkunen and C.A.Salgado, ``Constraints for the nuclear parton distributions from $Z$ and $W$ production at the LHC", {\it JHEP} {\bf 1103} (2011) 071, arXiv:101.5392 [hep-ph].
\bibitem{NRJloss} R.B.Neufeld, I.Vitev and B.W.Zhang, ``Toward a determination of the shortest radiation length in nature", {\it Phys.Lett. B} {\bf 704} (2011) 590-595. arXiv:1010.3708 [hep-ph].
\bibitem{CMS-PAS-HIN-12-008} CMS Collaboration, ``Study of Z boson production with 150~$\mu$b$^{-1}$ integrated PbPb luminosity at $\sqrt{s_{NN}} = 2.76$~TeV", {\it CMS PAS HIN-12-008} {\bf HIN-2012/008} (2012).
\bibitem{CMS_detector} CMS Collaboration, ``The CMS experiment at the CERN LHC", {\it JINST} 0803:S08004 2008.
\bibitem{Chatrchyan:2012vq} CMS Collaboration, ``Measurement of isolated photon production in $pp$ and PbPb collisions at $\sqrt{s_{NN}}=2.76$ TeV", {\it Phys.Lett. B} {\bf 710} (2012), 256-277, arXiv:1201.3093 [nucl-ex].
\bibitem{Reconstruct_GSF} W. Adam et al, `Reconstruction of Electrons with the Gaussian-Sum Filter in the CMS Tracker at the LHC", {\it J Phys. G: Nucl. Part. Phys.} {\bf 31} (2005) N9-N20, arXiv:physics/0306087.
\bibitem{CMS:2010bta} CMS Collaboration, ``Electron reconstruction and identification at sqrt(s) = 7 TeV", {\it CMS PAS EGM-10-004} {\bf EGM-2010/004} (2010).
\bibitem{Miller:2007ri} Miller, Michael L. and others, ``Glauber modeling in high energy nuclear collisions", {\it Ann. Rev. Nucl. Part. Sci.} {\bf 57} (2007) 205, arXiv:nucl-ex/0701025.
\end{thebibliography}
\end{document}